\documentclass{evn2004}
\usepackage{txfonts}
\usepackage{graphicx}
\begin{document}
   \title{First results of European VLBI radar observations \\ of space objects}

   \author{I.~Molotov\inst{1}, 
   G.~Tuccari\inst{2}, M.~Nechaeva\inst{3}, N.~Dugin\inst{3}, A.~Konovalenko\inst{4}
   I.~Falkovich\inst{4}, Y.~Gorshenkov\inst{5}, X. Liu\inst{6}, A.~Volvach\inst{7},
   V.~Agapov\inst{8}, A.~Pushkarev\inst{1}, V.~Titenko\inst{1,8}, S.~Buttacio\inst{2},
   V.~Rumyantsev\inst{7}, I.~Shmeld\inst{9}
          }

\institute{Central Atronomical Observatory, Pulkovo, Pulkovskoe sh. 65/1, 196140 St.-Petersburg, Russia
\and
Istituto di Radioastronomia, Contrada Renna Bassa, 96017 Noto, Italy
\and
Research Institute, B. Pecherskaya 25, 603950 N. Novgorod, Russia
\and
Institute of Radio Astronomy, Chervonopraporna str. 4, 310002 Kharkiv, Ukraine
\and
Special Research Bureau, MPEI, Krasnokazarmennaya 14, 111250 Moscow, Russia
\and
Urumqi Astronomical Observatory, NAO CAS, S. Beijing Road 40, 830011 Urumqi, China
\and
Crimean Astrophysical Observatory, 98409, Simeiz, Ukraine
\and
Keldysh Institute of Applied Mathematics, Miusskaja sq. 4, 125047               
\and
Institute of Astronomy, University of Latvia Boulevard Rainis 19, LV-1586 Riga, Latvia
}

   \abstract{Since 1999 we carried out seven trial VLBI radar experiments under LFVN project.
    The aim of this work is to adjust new research technique for investigating the Solar
    system bodies (planets, asteroids, space debris). It is planned to obtain the information
    on their movement parameters, proper rotation and structure of surface. The transmitter of
    Evpatoria RT-70 sounded the space objects. Array of Bear Lakes RT-64, Noto RT-32, Urumqi
    RT-25, Simeiz RT-22 received the echo-signals. The data were processed with NIRFI-3 Mk-2
    correlator in N. Novgorod, Russia and NRTV processor in Noto, Italy. The first results of
    these experiments are presented. 
   
  }
   \authorrunning{I. Molotov et al.}
   \titlerunning{First results of European VLBI radar observations of space objects}
   \maketitle

\section{Introduction}
In 1998, Evpatoria RT-70 was joined to the LFVN project (Molotov et al. \cite{molotov03}) that has the purpose to arrange the international VLBI cooperation with participation of former Soviet Union radio telescopes. The fully steerable 70 m dish is equipped with one from most powerful 6-cm band transmitter in Europe that was used for the deep spacecraft communications and the radar research of Mars, Venus and Mercury (Molotov \cite{molotov04}). Since 1999, the radar researches of the Solar system bodies were renewed using unique experience of LFVN team on the differential VLBI measurements of spacecraft trajectories (Alexeev et al. \cite{alexeev89}). The seven trial experiments were carried out to adjust the VLBI radar procedure, which joins the ``classic'' radar and differential VLBI measurements. Such combination can provide new scientific instrument to measure the variations of proper rotation of the Solar system bodies, determine their trajectories at Radio Reference Frame, to obtain the data on the object sizes and structure of surface. The radar system of Evpatoria RT-70 at Ukraine provided the sounding of space targets (Earth group planets, near-Earth asteroids and space debris objects) with help of 200~kW transmitter of continuous power at 5010~MHz. The reflected echo-signals were received at Bear Lakes RT-64, Noto RT-32, Urumqi RT-25, Simeiz RT-22 and 
some time in Torun RT-32, Shanghai RT-25, Kashima RT-34, Svetloe RT-32 and Kalyazin RT-64. 
At first, the Mk-2 and S2 VLBI terminals were used for recording the echoes (the recorded data were processed by NIRFI-3 Mk-2 correlator in N. Novgorod). Then, new specialized NRTV (near-real-time-VLBI) terminals and correlator were designed in Noto, Italy (Tuccari et al. \cite{tuccari02}). In common, the echoes of Mars, Venus and about 50 orbital objects were detected.

\section{Observations and processing}

The some results of the experiments VLBR03.1 are presented below. The session was carried out in the period of July 23 -- 29, 2003.
The scientific program included Venus, 2000 PH5 asteroid and a number of space debris objects at geostationary, highly-elliptical and
half-day circular orbits.

The first stage of processing includes the auto-correlation of the recorded tapes to detect the echo-signals from each
object on each receiving antenna.
The next stage is the cross-correlation processing of the transmitted signals and received echoes for each baseline
``transmitting antenna -- receiving antenna''. As a result of spectral analysis the frequency of doppler shift is measured.

\begin{figure}
   \centering
  \includegraphics{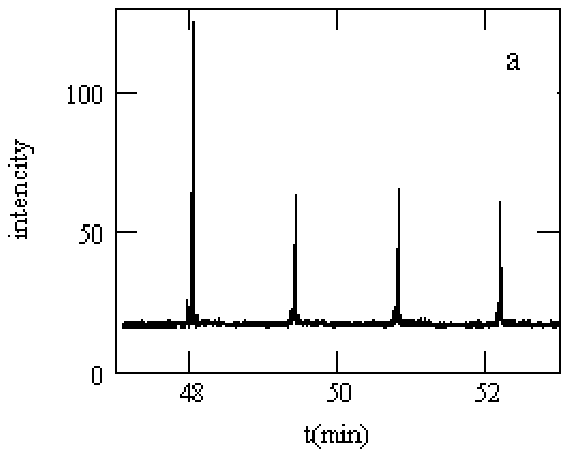}
  \includegraphics{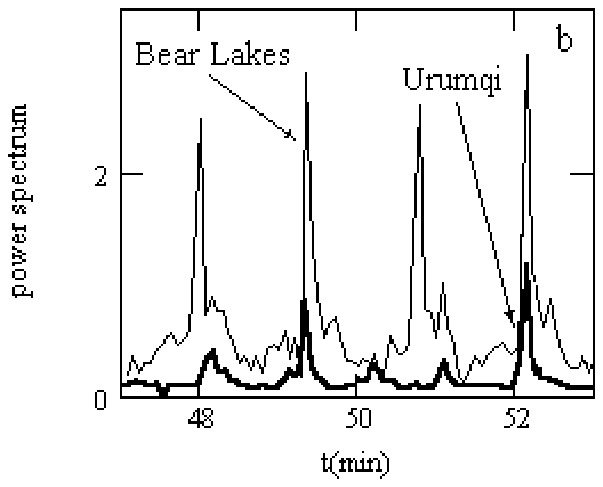}
    \caption{a) Recording of echo intensity of "Raduga-9"; time resolution dt=0.017 s; estimated period of rotation is 83 s;
    b) Dependence of spectral maximum on time (cross-spectrums for "Raduga-9" on Evpatoria-Bear Lakes and Evpatoria-Urumqi
    baselines), time resolution dt=4.26 s, real period of rotation is 166 s.
        \label{raduga}
        }
  \end{figure}

The recording of echo intensity at Bear Lakes RT-64 allows to evaluate the main period of the object rotation 
(Fig.~\ref{raduga}a). Fourier transformation of the single impulse of an object rotation gives the one-dimensional convolution of the
reflected region in wavelengths to estimate the sizes of the object. While the record of intensity is not performed at
other receiving sites, we are tracking the time evolution of maximum of cross-spectrum (``transmitting antenna --
receiving antenna''), that is equivalent to echo intensity changing (Fig.~\ref{raduga}b).

At final stage the measured fringe rate is analyzed and relative (echo - reference quasar) fringe rate value is
calculated.

\section{Results}

Currently, the rows of Doppler shift measurements were obtained for baselines Evpatoria -- Bear Lakes, Evpatoria -- Noto and Evpatoria -- Urumqi. In principle, it is enough to solve the coordinate goal using of object trajectory model (Alexeev et al. \cite{alexeev89}). The mean-root-square error is 0.096~Hz (corresponding to 3~mm/s rate), that is three times worse than apparatus mistake (1~mm/s rate).

The recording of intensity demonstrates the period of rotation for ``Raduga-9'' as 83~s, the time dependence of spectral
maximum on baseline Evpatoria -- Bear Lakes confirmed this result, while spectrum Evpatoria-Urumqi demonstrates in two time
longer period. This means that Raduga~9 has symmetric elements  (solar batteries) that are visible from Bear Lakes and
Urumqi under different angles.

\begin{figure}
   \centering
  \includegraphics{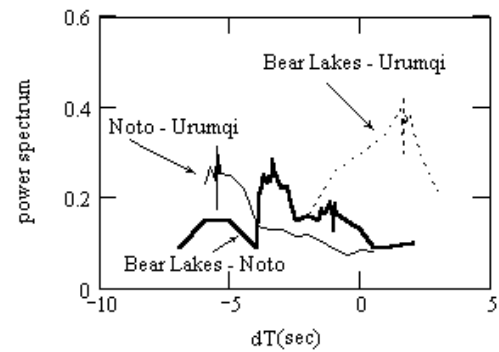}
    \caption{Time dependence of cross-spectrums maximums of "Cosmos-1366" at baselines Bear Lakes -- Noto, Bear Lakes --
     Urumqi and Noto -- Urumqi (25 July, 2003).    
        \label{cosmos}
        }
  \end{figure}

The VLBI fringes for the echoes of ``Cosmos-1366'' were received for the three baselines: Bear Lakes -- Noto -- Urumqi.
Analysis of time evolution of its spectral maximum (Fig.~\ref{cosmos}) at different baseline allows to get new information about proper 
movement of investigated object.
The shifts of cross-spectrum maximums were measured at relatively to moment of cross-spectrum maximum at
baseline Evpatoria (tramsmitted signal) -- Urumqi. These shifts equaled   to
$-3.35$~s for Bear Lakes-Noto, +1.65~s for Bear Lakes-Urumqi and $-5.5$~s for Noto-Urumqi baselines. Accordingly Bear Lakes
site, the echo signals ahead in Noto on 2.15~s and delayed in Urumqi on 5~s. It may be explained that scattering pattern
of ``Cosmos-1366'' is narrow (few degrees) and radiation maximum successively passed the receiving points during
rotation.  This fact allows to evaluate the direction of object  rotation axis.

\begin{acknowledgements}
This research was supported by INTAS 2001-0669 and RFBR 02-02-17568 grants. We thank the staff at the radio observatories Bear Lakes, Evpatoria, Noto, Urumqi, Simeiz that participated in the experiments and the optical observatories Pulkovo, Nauchny, Mayaki observations of that allowed improving the ephemerides of space objects. We thank James Dick and Phil Herridge for their
continuing support and tremendeous help with collecting of optical measurements by PIMS for studied objects.

\end{acknowledgements}


\begin{thebibliography}{}
  \bibitem[1989]{alexeev89} Alexeev, V. A., Altunin, V. I., 
     Antipenko, A. A., et al. 1989, Kosmicheskie Issledovania, 27, 765

  \bibitem[2003]{molotov03} Molotov, I., Kovalenko, A., 
     Samodurov, V., et al. 2003, Astronomical and Astrophysical Transactions, 22, 743
  
    \bibitem[2004]{molotov04} Molotov, E. P. 2004, Ground radio technique systems for control
    of spacecrafts (FIZMATLIT), 256
      
  \bibitem[2002]{tuccari02} Tuccari, G., Molotov, I., Buttacio, S., et al. 2002,
     in Proceedings of the 6th European VLBI Network Symposium, 
     eds. E. Ros, R. W. Porcas, A. P. Lobanov, and J. A. Zensus, 45
         
\end{thebibliography}
\end{document}